\documentclass[conference]{IEEEtran}
\IEEEoverridecommandlockouts
\usepackage{cite}
\usepackage{hyperref}
\usepackage{amsmath,amssymb,amsfonts}
\usepackage{bm}
\usepackage[normalem]{ulem} % https://stackoverflow.com/a/51134015

\newcommand{\rtsixty}{\text{RT}_{60}}
\DeclareMathOperator{\STFT}{STFT}
\def\C{\mathbb{C}}

\def\thineq{\hspace{-.1em}=\hspace{-.1em}}

\usepackage{algorithmic}
\usepackage{graphicx}
\usepackage{textcomp}
\usepackage{xcolor}
\usepackage{comment}
\usepackage{siunitx}
\usepackage{multirow}
\usepackage{pifont}% http://ctan.org/pkg/pifont
\newcommand{\cmark}{\ding{51}}%
\newcommand{\xmark}{\ding{55}}%

\def\BibTeX{{\rm B\kern-.05em{\sc i\kern-.025em b}\kern-.08em
T\kern-.1667em\lower.7ex\hbox{E}\kern-.125emX}}

\newcommand{\louis}{}
\newcommand{\louiserase}[1]{}

\newcommand{\mathieucomment}[1]{}
\newcommand{\louiscomment}[1]{}
\newcommand{\gaelcomment}[1]{}

\makeatletter
\def\ps@IEEEtitlepagestyle{%
  \def\@oddfoot{\mycopyrightnotice}%
  \def\@evenfoot{}%
}
\def\mycopyrightnotice{%
\begin{minipage}{\textwidth}
  \centering
  \footnotesize \copyright 2025 IEEE. Personal use of this material is permitted. Permission from IEEE must be obtained for all other uses, in any current or future media, including reprinting/republishing this material for advertising or promotional purposes, creating new collective works, for resale or redistribution to servers or lists, or reuse of any copyrighted component of this work in other works.
  \end{minipage}
}
\makeatother

\begin{document}

\title{
  % Speech dereverberation supervised by Polack's Model\\
  % Deep speech dereverberation supervised only by reverberation time \\
  % \mathieu{$\Phi$-WSD: Physically weakly-supervised speech dereverberation}
  A Hybrid Model for Weakly-Supervised\\ Speech Dereverberation
% {\footnotesize \textsuperscript{*}Note: Sub-titles are not captured for https://ieeexplore.ieee.org  and
% should not be used}
  \thanks{
  This work was funded by the European Union (ERC, HI-Audio,
101052978). Views and opinions expressed are however those
of the author(s) only and do not necessarily reflect those of the
European Union or the European Research Council. Neither
the European Union nor the granting authority can be held responsible for them. This work was performed using HPC resources from GENCI–IDRIS (Grant 2024-AD011014072R1).}
}

\author{
    \IEEEauthorblockN{\textit{
    Louis Bahrman\IEEEauthorrefmark{1}, 
    Mathieu Fontaine\IEEEauthorrefmark{1},
    Gaël Richard\IEEEauthorrefmark{1}
    }
}
\vspace{0.5cm}
\IEEEauthorblockA{
    \IEEEauthorrefmark{1}LTCI, T\'el\'ecom Paris, Institut Polytechnique de Paris, Palaiseau, France\ %\hspace{0.2cm}
    }
}

\maketitle

\begin{abstract}

This paper introduces a new training strategy to improve speech dereverberation systems using minimal acoustic information and reverberant (wet) speech. Most existing algorithms rely on paired dry/wet data, which is difficult to obtain, or on target metrics that may not adequately capture reverberation characteristics and can lead to poor results on non-target metrics.
Our approach uses limited acoustic information, like the reverberation time (RT60), to train a dereverberation system. The system's output is resynthesized using a generated room impulse response and compared with the original reverberant speech, providing a novel reverberation matching loss replacing the standard target metrics. During inference, only the trained dereverberation model is used.
Experimental results demonstrate that our method achieves more consistent performance across various objective metrics used in speech dereverberation than the state-of-the-art.

\louiscomment{
Most deep-learning-based approaches for monaural dereverberation require dry signals at the training stage. 
%  Discriminative deep-learning-based approaches for speech dereverberation often requires pairs of reverberant and dry signals
% are often supervised by pairs of reverberant and dry signals. 
  However, such dry signals have to be recorded in anechoic conditions and are difficult to obtain in practice. 
  In this work, we address this problem by introducing a novel training framework which leverages a conventional reverberation model to supervise a learnt dereverberation model, requiring only pairs of reverberant signals and reverberation times for its hybrid weakly-supervised training.
  At inference, the reverberation time is not needed, 
  % In this work, we address this problem by using a novel training procedure, which leverages Polack's room model to supervise a dereverberation model, requiring only the reverberation time.
  % Few approaches leveraging only reverberant signals have been proposed, and either need overdetermined mixtures or metrics-based supervision. 
and our approach yields better and more consistent results than training strategies relying on a metric.
}
\end{abstract}


\section{Introduction}

%%%%%%%%%%%%%% INTRO V2 %%%%%%%%%%%%%

Acoustic signals captured in closed rooms are affected by reflections from room walls and diffraction from obstacles on its path, in a process coined as reverberation.
These effects may not be desirable in speech recordings as they lower speech intelligibility \cite{10.1121/1.392224}.
This justifies the need for dereverberation methods to mitigate the reverberation phenomenon in speech-related tasks such as speech enhancement and automatic speech recognition \cite{yoshioka_making_2012}. 
Dereverberation task has been historically solved by using statistical signal-processing methods \cite{nakatani_speech_2010,lebart_new_2001,Gaubitch2010_dereverb_lpc}. 
%\mathieucomment{tu peux égayer la liste si tu le souhaites ;) . Tu peux regarded dans la thèse d'Arthur http://www.atiam.ircam.fr/Archives/Stages1314/BELHOMME_Arthur_Rapport1314.pdf} \louiscomment{Merci, mais j'ai retrouvé des approaches LPC}
% \mathieustrike{but state-of-the-art approaches for dereverberation are based on deep learning, and their performance is highly dependent on the amount of data they are trained on.}
The nonlinearity of the task naturally calls for deep neural networks (DNNs) extensions requiring in practice a large amount of annotated data and learning strategies.
These learning-based approaches can be supervised in various ways. 

\begin{figure}[t]
    \centering
    \includegraphics[width=.9\linewidth]{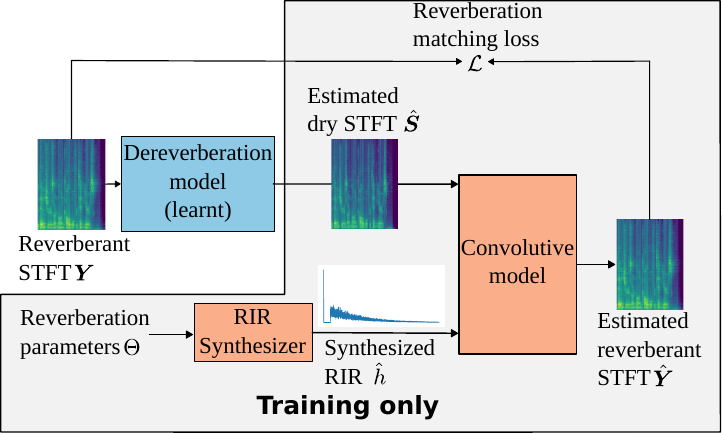}
    \louis{\vspace{-0.5em}}
    \caption{Overview of the proposed method}
    \label{fig:overview}
    \vspace{-1em}
\end{figure}

Best-performing discriminative approaches as TFGridNet~\cite{wang_tf-gridnet_2023} learn to predict a dry signal from a reverberant one decomposing the time-frequency signal into time and spectral subband modules and hence require paired dry/wet data. FullSubNet \cite{hao_fullsubnet_2021} aims to estimate a Complex Ratio Mask (cRM) for retrieving the dry signal.
However, these techniques require generating a large amount of paired data and may lack robustness if the test data significantly differs from the training dataset.
This lack of paired data has motivated the development of new approaches which can dereverberate using unpaired signals only as in
%To alleviate the lack of paired of reverberant and dry data, 
Cycle-consistent Generative Adversarial Networks (GANs) \cite{muckenhirnCycleGANbasedUnpairedSpeech2022a, yuCycleGANbasedNonparallelSpeech2021}.% \mathieu{can dereverberate only using unpaired signals.}

% Even if they do not require reverberant speech samples in their training set, generative models still rely on datasets of dry speech signals. 
On the other hand, generative models like 
variational autoencoders~\cite{fangVariationalAutoencoderSpeech2021,leglaiveRecurrentVariationalAutoencoder2020} or diffusion models~\cite{lemercierAnalysingDiffusionbasedGenerative2022} 
are used to learn the prior dry signals without having access to any reverberant signal at training.
Although these models require less supervision, they do not solve the problem of data scarcity, since dry speech data is harder to obtain than reverberant speech data. 

Few approaches are designed to require only reverberant signals at training time. 
The current best-performing model for dereverberation supervised only by reverberant signals is MetricGAN-U \cite{fuMetricGANUUnsupervisedSpeech2022}. Its training framework is based on a GAN, where the discriminator is trained to mimic the behaviour of a target metric, and the generator to optimize its performance with respect to this evaluation metric. 
It has been successfully applied on the dereverberation task, using Speech to Reverberant Modulation energy Ratio (SRMR) \cite{falkNonIntrusiveQualityIntelligibility2010} as a target metric to be optimized.
 
Besides, both supervised and unsupervised approaches for dereverberation have been improved by leveraging reverberation models. 
Such approaches can be considered as hybrid deep learning, in the sense that they combine DNN priors with statistical models or signal-based representations of the reverberation.
Indeed, reverberation has been classically represented as a convolutive distortion and approaches
have been developed to concurrently estimate the convolutive model and the dry signal~\cite{vincent_audio_2018, nakatani_speech_2010, lebart_new_2001}. 
The reverberation model can be implicitly modelled, or explicitly used. 
A popular choice to implicitly model reverberation is the Convolutive Transfer Function (CTF) approximation, which considers reverberation as a subband filtering process. It has been used in the weighted prediction error (WPE) method~\cite{nakatani_speech_2010} and its neural enhancements \cite{kinoshita_neural_2017} or with diffusion models \cite{saito_unsupervised_2023} and variational approaches \cite{rvae-em}. 
An observation model based on CTF has been used in conjunction with discriminative approaches under the name Forward Convolutive Prediction (FCP) \cite{wang_convolutive_2021}.
FCP has even been used for unsupervised learnt dereverberation in USDNet \cite{wangUSDnetUnsupervisedSpeech2024a}. USDNet is powerful in a multichannel setting but shows only subpar performance in a monaural setting.
\louiscomment{Clarified that USDNet is powerful, not CTF}

The convolutive model can also be explicitly modelled.
Recent work shows that models designed to only dereverberate are also able to explicitly model reverberation \cite{bahrman24_interspeech} and that it is possible to constrain a diffusion posterior sampling to match acoustic properties \cite{molinerBUDDySingleChannelBlind2024}.
Other methods even assume that some reverberation properties are available at inference \cite{wu_reverberation-time-aware_2017,li_composite_2023}, or even the full room impulse response (RIR) which uniquely characterizes the reverberation process \cite{lemercier_diffusion_posterior_sampling_2023}. 
This has been made possible by the recent advances in blind room acoustic parameters estimation
\cite{prego_2015_blind}\louis{,\cite{MackSingleChannelBlindDirecttoReverberation2020}}.
% \louis{, where room parameters are estimated from reverberant speech only, requiring no dry data}\cite{prego_2015_blind}\louis{,\cite{MackSingleChannelBlindDirecttoReverberation2020}}.

So far, MetricGAN-U remains the best approach for dereverberation supervised by reverberant signals\louiserase{.}\louis{, outperforming unsupervised approaches such as WPE.}
We qualify this approach as a \textit{metrics-based weak supervision} since it requires a nonintrusive metric to compute its training target. 
However, metrics-based supervision is known to potentially be detrimental to performance regarding other criteria~\cite{deoliveira24_interspeech}.

In this article, we propose to alleviate this issue by introducing a novel hybrid weak supervision framework for dereverberation, called \textit{reverberation-based weak supervision}.
% We train a deep neural network such that its dry speech estimate on which a classical reverberation model is applied, matches its reverberant input signal
We train a deep neural network to predict a dry estimate from a reverberant signal, such that a reverberation model applied to the dry estimate matches its reverberant input. 
We show that reverberation-based weak supervision performs better than metrics-based weak supervision on various objective measures. 
For reproducibility purposes and to help future research, we publicly distribute examples, code and pretrained models \footnote{\url{https://louis-bahrman.github.io/Hybrid-WSSD/}}.

\section{Reverberation model}

\subsection{Late reverberation and mixing time}
Assuming fixed source and microphone positions and no additive noise, a monaural reverberant signal $y$ can be represented as a convolution between a dry signal $s$ and the room impulse response (RIR) $h$ between the source and the microphone:
\begin{equation}
    y(n) = (s \star h)(n), \label{eq:time_domain_convolution}
\end{equation}
where $n$ denotes the time index and $\star$ the convolution operator. 
The RIR $h$ can be divided into three parts: the direct path corresponds to its first peak $h_d$ followed by  early echoes $h_e$ and, after the mixing time, late reverberation $h_l$:
\louiscomment{Facultative equation}
\begin{equation}
    h= h_d + h_e + h_l,
\end{equation}
The support of $h_d$, $h_e$ and $h_l$ are disjoint, and several definitions for the mixing time $n_m$ have been proposed. In ~\cite{blesser2001synthesisOfReverbViewPoints}, the mixing time in samples is defined using statistical properties of ergodic rooms, as a multiple of  the mean free path~\cite{joyceSabineReverberationTime1975}:
\begin{equation}
    n_m = \frac{4V f_s}{c A}, \label{eq:mixing_time}
\end{equation}
where $V, f_s, c$ and $A$ are respectively the room volume, \louiserase{the }sampling frequency, \louiserase{the }speed of sound and \louiserase{the }area of the walls.

\subsection{Polack's late reverberation model }
% Unlike other approaches, we consider late reverberation as a convolutive (not additive) model.
A simple yet powerful model for late reflections is \louiserase{the so-called }Polack's model~\cite{polackTransmissionEnergieSonore1988}.
This model states that the late reverberation $h_l$ is a realization of an exponentially decaying white noise:
\begin{equation}
h_l(n) = b(n) e^{-(n+3 n_m)/ \tau}, 
\end{equation}
where $b(n) \sim \mathcal{N}(0, \sigma^2)$ is a centered Gaussian distribution, and $\tau$ depends on the reverberation time $\rtsixty$ and the sampling frequency $f_s$ as\louis{:}
\begin{equation}
    \tau=\frac{\rtsixty f_s}{3 \ln(10) }.
\end{equation}

\subsection{Convolution in Time-Frequency domain}
The time-invariant linear system of Eq. \eqref{eq:time_domain_convolution} can be formulated in the short-time Fourier transform (STFT) domain as interband and interframe convolution~\cite{avargel_system_2007}:
\begin{equation}
Y_{f,t}=\sum_{f^{\prime}=0}^{F-1}\sum_{t^{\prime}=0}^{\min(t; T_h)} \mathcal{H}_{f,f^{\prime},t^{\prime}} S_{f^{\prime},t-t^{\prime}}, \label{eq:full_convolution}
\end{equation}
\louiscomment{En fait il faut utiliser à la fois $T_h$ et $N_h$, car la longurur de $\mathcal{H}$ et de $h$ ne sont pas les mêmes}
where $\bm{Y} \triangleq \{Y_{f,t}\}_{f,t=0}^{F-1, T_y-1} \in \mathbb{C}^{F \times T_y}$ are the STFT coefficients of the reverberant signal at frequency $f\thineq 0,\dots,F-1$ and time $t\thineq 0,\dots,T_y-1$, 
$\mathcal{H} \triangleq \{\mathcal{H}_{f, f^\prime, t}\}_{f,f^\prime, t=0}^{F-1,F-1,T_h-1} \in \C^{F \times F \times T_h}$ is a tridimensional representation of the RIR 
and $\bm{S} \triangleq \{S_{f,t}\}_{f,t=0}^{F-1, T_s-1}\in \C^{F\times T_s}$ is the $\STFT$ of the dry signal.
As shown in~\cite{avargel_system_2007}, $\mathcal{H}$ can be obtained in closed form from the RIR $h \in \mathbb{R}^{N_h}$ as: 
% \begin{equation}
%     \frac{1}{F}
%     \sum_{m=-N+1}^{N-1} h((t-t')L - m) 
%     \sum_{n=0}^{N-1}
%     w_s(n + m) e^{j2\pi f' m/F}
%     w_a(n) e^{j2 \pi (f'-f)n /F}
% \end{equation}
\begin{equation}
    \mathcal{H}_{f, f', t'}
    =
    \sum_{m=-N+1}^{N-1} h(t'L - m) W_{f,f^\prime}(m), \label{eq:big_H}
\end{equation}
where $N$ is the $\STFT$ window length, $L$ the hop-size and
\begin{equation}
     W_{f,f^\prime}(m)=\frac{1}{F}\sum_{n=0}^{N-1}
    w_s(n + m)w_a(n) e^{\frac{j2\pi (f^\prime (n+m) - fn)}{F}} \label{eq:big_window} 
\end{equation}
 with $w_s, w_a$ the synthesis and analysis windows respectively.

\section{Proposed method}

\subsection{Overview}

We propose to supervise the training of a dereverberation deep neural network (DNN) using a conventional reverberation model. 
The general training procedure is as follows. 
Given a reverberant signal $\bm{Y}$ as in the previous section, the DNN outputs a dry signal estimate $\hat{\bm{S}} \triangleq \{\hat{S}_{f,t}\}_{f,t=0}^{F-1, T_s-1}\in \C^{F\times T_s}$.
In parallel, a reverberation model $\mathcal{R}$ synthesizes an approximated RIR $\hat{h}$ from a few reverberation model parameters $\Theta \triangleq \{\rtsixty,\sigma, V,A\}$.
Both the estimated dry $\STFT$ $\hat{\bm{S}}$ and the synthesized RIR $\hat{h} \in \mathbb{R}^{N_h}$ are convolved in a cross-band convolutive model $\mathcal{C}$ (see Eq.~\eqref{eq:cb_convol_model}), to compute an estimate of the reverberant spectrogram $\hat{\bm{Y}}$.
The standard dereverberation loss requiring pairs of dry and wet signals is replaced by a reverberation matching loss $\mathcal{L}$, computing the distance between the estimated and ground-truth reverberant spectrograms $\hat{\bm{Y}}$ and $\bm{Y}$.
A diagram of the training procedure is shown in Fig.~\ref{fig:overview}.
Because the RIR synthesis and convolutive model are not parametric, they do not need to be trained. These blocks are discarded at inference, and only the DNN is used. Hence, the number of parameters, as well as the computational complexity and memory footprint are the same
as for the original model.

\subsection{RIR synthesizer}

The RIR synthesizer aims at synthesizing an RIR for which the late reverberation $h_l$ matches Polack's model, and the direct path $h_d$ is a peak of amplitude $1$. 
To better match Polack's model with our data distribution without changing its energy distribution,
and based on preliminary experiments, we decided to synthesize an RIR using the absolute value of the Gaussian distribution used in Polack's model.
According to the mean free path property, the direct path peak should be on average positioned at the sample corresponding to the mean free path of the room $n_m$. As stated in~\cite{blesser2001synthesisOfReverbViewPoints}, the mixing time, corresponding to the beginning of the late reverberation $h_l$, is set at $3$ times the mean free path. 
However, to better align the dry and reverberant signals, we discard the RIR samples before the first peak. 
Hence, the synthetic RIR becomes:
\begin{equation}
    \mathcal{R}(\Theta)(n) = \begin{cases}
        | b(n) | e^{-\frac{3 \ln(10)}{\rtsixty f_s} n } & \text{if } n > 2n_m\\
        1 & \text{if } n = 0\\
        0 & \text{otherwise},   
    \end{cases}
\end{equation}
where $b(n)$ is drawn from a Gaussian distribution $\mathcal{N}(0, \sigma^2)$, and $n_m$ corresponds to Eq.~\eqref{eq:mixing_time}.
\louis{In this model, at fixed $\rtsixty$, $\sigma^2$ is proportional to the inverse of the direct-to-reverberant ratio (DRR), which has been proven to have great influence on dereverberation performance \cite{habetsLateReverberantSpectral2009}.}
% and can be estimated from wet speech only~\cite{mackSingleChannelBlindDirecttoReverberation2020}.}

\subsection{Convolutive model and reverberation matching loss}

To better backpropagate the training gradient to the dereverberation model whose output might be in the time-frequency plane, we consider a time-frequency cross-band convolutive model and reverberation matching loss. 
Given $\hat{h} = \mathcal{R}(\Theta)$, and $\hat{\bm{S}}$ the dry speech estimate outputted by the DNN, we define the time-frequency convolutive model as:
% \begin{align}
%     \hat{\bm{Y}}_{f,t} &= \mathcal{C}(\hat{\bm{S}}, \hat{h}) \\
%     &= 
%     \sum_{f^{\prime}=f-F^{\prime}}^{f+F^{\prime}}\sum_{t^{\prime}=-\infty}^{\infty} S_{f^{\prime},t-t^{\prime}} \label{eq:cb_convol_model}
%     \sum_{m=-N+1}^{N-1} h(t'L - m) W_{f,f'}(m), 
% \end{align}

\begin{align}
    \hat{\bm{Y}}_{f,t} \triangleq \mathcal{C}(\hat{\bm{S}}, \hat{h}) 
    = 
    \sum_{f^{\prime}=f-F^{\prime}}^{f+F^{\prime}}\sum_{t^{\prime}=0}^{\min(t; T_h)} \hat{\mathcal{H}}_{f, f', t'}\hat{S}_{f^{\prime},t-t^{\prime}} \label{eq:cb_convol_model}
    , 
\end{align}
with $\hat{\mathcal{H}}_{f, f', t'} \triangleq \sum_{m=-N+1}^{N-1} \hat{h}(t'L - m) W_{f,f'}(m)$ and the notations in Eq.~\eqref{eq:cb_convol_model} coinciding to those of Eq.~(\ref{eq:full_convolution}-\ref{eq:big_window}).
Based on \cite{avargel_system_2007}, we set the number of crossbands $F'$ to $4$.

Our reverberation-matching loss corresponds to the commonly used mean-squared error estimator for the deconvolution problem. Since this problem can be ill-posed for low-amplitude signals, a regularization term matching the log-magnitudes of the reverberant estimate and ground truth is added, and the model training loss is, with $\lambda=\gamma=1${ as 
in}~\cite{schwarMultiScaleSpectralLoss2023}:
\begin{equation}
 \mathcal{L} = \sum_{f,t} \left[\lvert \hat{Y}_{f,t}  - Y_{f,t} \rvert^2 + \lambda \left| \log\left(\frac{1 + \gamma \lvert \hat{Y}_{f,t}\rvert}{1 + \gamma \lvert Y_{f,t} \rvert} \right) \right|^2 \right]
\end{equation}

\section{Experiments}

We compare our proposed "reverberation-based weak supervision" with a baseline "metrics-based weak supervision" as implemented in MetricGAN-U.
\subsection{DNN variants}

We assess several variants of our method with FullSubNet (FSN)~\cite{hao_fullsubnet_2021}.
The ability of FullSubNet to process complex STFT representations both in the full-band and subband directions is required
to be paired with our proposed cross-band convolutive model and reverberation matching loss. 
It has also been proven to be able to jointly model physical properties of a room and dry speech \cite{bahrman24_interspeech}, and to be paired with reverberation-informed training strategies \cite{zhou_2023_reverb_time_shortening}.
We also consider the baseline BiLSTM model \cite{weningerSpeechEnhancementLSTM2015} used as a generator in MetricGAN-U. 
This model is much simpler as it only allows to processing STFT magnitude masks, 
and will serve as an indicator for the behaviour of our proposed loss with a less expressive model. 

\subsection{Supervision variants}
We considered several supervision variants classified as weak supervision and strong supervision.
% \par\noindent \textbf{Weak supervision (WS):}
\par\noindent \underline{Weak supervision (WS):}
WS variants include using Polack's model with either 
\begin{itemize}
    \item $\Theta \triangleq \{\rtsixty,\sigma, V,A\}$: all the parameters, including those used to estimate the mixing time. 
    \item $\{\rtsixty, \sigma \}$:  a fixed mixing time set as $20$~ms after the peak, corresponding to the mean of all mixing times in the training dataset.
    \item  $\{ \rtsixty \}$:  a fixed mixing time at $20$~ms and a median value of Polack's variance $\sigma = 0.02$ over the training dataset.
    This is the least-supervised model and is motivated by realistic scenarios where only the reverberation time can be computed from a reverberant signal~\cite{prego_2015_blind}.
\end{itemize}
% \par\noindent \textbf{Strong Supervision:}
\par\noindent \underline{Strong Supervision:}
Those variants include using more oracle information such as    
\begin{itemize}
    \item the exact RIR $h$ as an oracle RIR synthesis model. This variant should be considered as an upper bound for our proposed reverberation-based weak supervision's performance, as it is equivalent to having access to pairs of dry and reverberant signals as supervision.
    \item each model's original paired training loss as supervision. BiLSTM is trained using the mean squared error between dry and dereverberated magnitude spectra. FSN is trained to minimize the euclidean distance between its estimated and the ground-truth ideal complex ratio mask (cRM).
\end{itemize}
We also consider MetricGAN-U's metrics-based weak supervision as a baseline. It corresponds to the BiLSTM model trained with the weak supervision of the SRMR metric.

\subsection{Miscellaneous configurations}
As in the original FullSubNet, 49151 sample excerpts (around 3 s at 16 kHz) reverberant audios are processed in the STFT domain using a 512-sample Hann window with an overlap of $50~\%$. 
We use a learning rate decay based on the \louiserase{rereverberation}\louis{training} loss on a validation set, and early stopping based on the SISDR metric on a validation set. 

\subsection{Dataset}
Similarly to \cite{bahrman24_interspeech, wang_tf-gridnet_2023}, we simulated a training dataset by dynamically convolving dry speech signals with simulated RIRs. 
The dry speech signals are randomly sampled from the close-talking
microphone recordings in the WSJ1 dataset~\cite{wsj1}. The training
set is composed of a total of 73 hours of recordings split into 
60307 audio excerpts. The simulated RIR dataset consists of
32,000 RIRs drawn from 2000 rooms simulated using the image source method implemented in the pyroomacoustics library \cite{scheibler_pyroomacoustics_2018}.
Room dimensions and RT60 are uniformly
sampled in the respective ranges of $[5,10] \times [5,10] \times [2.5,4]~\text{m}^3$, and $[0.2,1.0]$~s.
The source-microphone distance is uniformly distributed in $[0.75, 2.5]$~m, 
and both source and microphone are at least $50$~cm from the
walls.
At training time, we use a dynamic mixing procedure
consisting of randomly selecting a dry signal and RIR pair. 
In order to align the dry signal target and the direct-path, the samples before the direct path are discarded and it is normalised (so that the direct-path is of amplitude 1). This does not change
the RIR distribution and compensates for the delay induced by
the direct-path to match the RIR synthesis procedure.

\section{Results and discussion}

\begin{table}[t]
  \centering
  \caption{Dereverberation scores $\pm$ standard deviation\\
  (for each metric, the higher the better)
  }
  \vspace{-.5em}
  \sisetup{
detect-weight, %
mode=text, %
tight-spacing=true,
round-mode=places,
round-precision=2,
table-format=1.2,
table-number-alignment=center
}
\resizebox{\linewidth}{!}{
\setlength{\tabcolsep}{3pt}
  \begin{tabular}{l|c |r|*{1}{S[round-precision=1,table-format=2.1]@{\,\( \pm \)\,}S[round-precision=1,table-format=1.1]S@{\,\( \pm \)\,}SS@{\,\( \pm \)\,}SS[round-precision=1,table-format=2.1]@{\,\( \pm \)\,}S[round-precision=1,table-format=1.1]S}}
    Model & \multicolumn{1}{c|}{WS?} & \multicolumn{1}{c|}{Supervision} & \multicolumn{2}{c}{SISDR} & \multicolumn{2}{c}{ESTOI} & \multicolumn{2}{c}{WB-PESQ} & \multicolumn{2}{c}{SRMR}\\
    \hline
    \multirow{5}{*}{FSN} 
    & \multirow{2}{*}{\xmark}& cRM & 5.611 & 3.941 & 0.839 & 0.09448 & 2.549 & 0.6773 & 8.239 & 3.509\\
     & & $h$ & 4.28 & 3.965 & 0.7682 & 0.1196 & 2.028 & 0.686 & 7.801 & 3.103\\
     \cline{2-11}
     & \multirow{3}{*}{\cmark} & $\Theta$ & 0.9851 & 2.54 & \bfseries \num{0.7078} & \bfseries \num{0.1449} & \bfseries \num{1.804} & \bfseries \num{0.6966} & 6.908 & 2.757\\
     & & $\{\rtsixty,\sigma\}$ & 1.1 & 2.522 & 0.702 & 0.1428 & 1.782 & 0.6856 & 7.045 & 2.812 \\
     & & $\{\rtsixty \}$ &  \bfseries \num{2.904} & \bfseries \num{3.388} & 0.7072 & 0.1464& 1.781 & 0.7072 & 6.914 & 2.783\\
     \hline
     \multirow{5}{*}{BiLSTM} 
     & \multirow{2}{*}{\xmark} & $\lvert S_{f,t} \rvert^2,\forall f,t $
     & 1.285 & 4.187 & 0.7768 & 0.1205 & 2.25 & 0.7924 & 7.882 & 3.032 \\
    & & $h$ & 0.1231 & 4.081& 0.7038 & 0.1501 & 1.797 & 0.6992 & 7.163 & 2.663 \\
    \cline{2-11}
    & \multirow{3}{*}{\cmark} & $\Theta$ & 0.8073 & 3.984& 0.6973 & 0.154 & 1.805 & 0.7441 & 6.933 & 2.736 \\
     & & $\{\rtsixty,\sigma\}$ & 0.7036 & 4.007 & 0.6956 & 0.1517 & 1.778 & 0.7188 & 6.846 & 2.739\\
     & &  $\{\rtsixty\}$ & \bfseries \num{1.609} & \bfseries \num{3.667} & \bfseries \num{0.7081} & \bfseries \num{0.1499} & \bfseries \num{1.837} & \bfseries \num{0.748} & 6.865 & 2.795\\
    \hline
    BiLSTM & \cmark & SRMR \cite{fuMetricGANUUnsupervisedSpeech2022} & -1.4891304969787598 & 3.3998758792877197& 0.6359751 & 0.18153876066207886& 1.7828631401062012 & 0.7352843284606934 & \bfseries \num{10.875326156616211} & \bfseries \num{4.2486042976379395} \\
    \hline
    \multicolumn{3}{c|}{Reverberant} & -1.32 & 3.413 &0.6847 & 0.1614 & 1.753 & 0.7435 & 6.917 & 2.861 \\
  \end{tabular}
  }
  \label{tab:results_v1}
  \vspace{-2em}
\end{table}

We evaluate the performance of our proposed reverberation-based weak supervision for the dereverberation task on unseen speakers from WSJ1 (Hub and Spokes S1 to S4) and rooms. 
The performance is evaluated using the Scale-Invariant Signal-to-Noise ratio (SISDR) \cite{rouxSDRHalfbakedWell2019}, Extended Short-Time Objective Intelligibility (ESTOI) \cite{stoi}, Wide-Band Perceptual Evaluation of Speech Quality (WB-PESQ)\cite{wb_pesq} and SRMR \cite{falkNonIntrusiveQualityIntelligibility2010} metrics.
To assert the statistical significance of our result analysis despite the high measured variances, we opted for a non-parametric Wilcoxon test with a significance level of $0.001$ for the null hypothesis to be rejected.
The results are presented in Table \ref{tab:results_v1}. 
The line denoted "Reverberant" corresponds to unprocessed signals and
the best significant weak supervision variant for each metric and dereverberation model is highlighted.
% We are better
All of the proposed methods show an improvement of the SISDR, ESTOI and WB-PESQ metrics, meaning that they can successfully dereverberate speech. 
% Metrics-based is bad
The baseline (BiLSTM + SRMR) excels in terms of SRMR, but this performance comes at the cost of its SISDR and ESTOI results, which are degraded compared to the reverberant input. 
This result confirms the main drawback of metrics-based weak supervision, in the sense that it tends to 
solely optimize the metric it is trained on, disregarding the others. 
Indeed, all our proposed methods perform better than the baseline on all other metrics than SRMR. 
This demonstrates the superiority of reverberation-based weak supervision over metrics-based weak supervision.
% Weak vs Strong
The best-performing method FullSubNet benefits from strong supervision, both when trained on its original complex masking loss or using the oracle RIR. 
On the other hand, 
the less-complex BiLSTM widely benefits from weak supervision, and performs better in terms of SISDR when weakly supervised by Polack's model than when it has access to the ground-truth RIR $h$. For this model, the weakest supervision by $\rtsixty$ yields not only superior results to other weak-supervision variants 
for all metrics except SRMR,
but even improves the model's SISDR performance above its original supervision based on magnitude spectra.
This is due to the BiLSTM's design, which is meant only for a spectral magnitude masking loss, without alleviating the STFT phase.
Hence, when the estimated dry signal is reverberated using a ground-truth RIR $h$, the estimated reverberant STFT phase is perturbed to a large extent, whereas reverberation by Polack's model yields a phase that is closer to the complex circular Gaussian model at the core of BiLSTM's design. 
Another noticeable result occurs for both models in the reverberation-based weakly supervision by Polack's model.
% Between Weak supervision approaches
Comparing reverberation-weak supervision approaches, we remark that 
they perform better in terms of SISDR when having no access to the acoustic parameters used to estimate the mixing time and Polack's model $\sigma$.
\louiserase{Observation of the training loss shows that both FSN and BiLSTM converge to distinct optima when trained using the $\rtsixty$ information only and using all $\{\rtsixty,\sigma, V, A \}$ as weak supervision.}
\louis{Hence, fixing $\sigma, V$ and $A$ is equivalent to making the DRR only dependant on the $\rtsixty$, which can be easily computed from reverberant speech~\cite{prego_2015_blind}, and seems to regularize our proposed training procedure for dereverberation %.}
 when evaluated with synthetic RIRs.}
% \louis{Indeed, fixing $\sigma$ is equivalent to making the DRR only dependant on the $\rtsixty$, which seems coherent with the data generation procedure}
\louiserase{Weaker reverberation based supervision techniques regularize better the dereverberation training.}

% \louiscomment{Scoresobjectifs perceptuel, take home message: 
% Perceptive tests
% }
% I think that difference between ISM's model and real RIR is smaller than difference between Polack and ISM. Seems to be able to generalize well. If I have time Real RIR dataset for testing (2h to implement the dataset + 1h30 to get the results)

\section{Conclusion}

We have proposed a novel approach for weakly-supervised speech dereverberation, by training a deep neural network to predict a dry estimate from a reverberant signal, such that a reverberation model applied on
the dry estimate matches its reverberant input.
Results demonstrate the superiority of our reverberation-based weak supervision over metrics-based weak supervision.
This method opens the path to a variety of \louis{dereverberation}\louiserase{derereverberation} techniques for data-scarce scenarios and various signals such as music.
Future work will be dedicated to leveraging a more powerful RIR synthesis model that can estimate the $\rtsixty$ from reverberant signals only{, and to extending this work to weakly-supervised generative approaches for dereveberation to better model the probabilistic RIR model.}

\bibliographystyle{ieeetran}
\bibliography{biblio}

\end{document}